# Sonnenstrom aus Plastik

Vielversprechende Solarzellen aus konjugierten Molekülen und Polymeren
Carsten Deibel und Vladimir Dyakonov

**Durch den Erfolg organischer Leuchtdioden beflügelt, entwickeln sich auch organische Solarzellen zu interessanten Anwendungen, welche das Potential haben, die Anwendungsbereiche anorganischer Systeme zu ergänzen und erweitern.**

Günstige, effiziente und flexible Solarzellen, die mittels eines Druckprozesses bei Raumtemperatur hergestellt werden können — nur eine schöne Zukunftsvision? Die aktuellen Entwicklungen im Bereich der organischen Photovoltaik zielen genau in diese Richtung. Viele organische Materialien können bei niedrigen Temperaturen verarbeitet werden. Daher können auch flexible Plastikfolien, die nicht so temperaturstabil sind, als Substrate genutzt werden (s. Abb. 1). Somit ist die schon etablierte Rolle-zu-Rolle Prozessierung, welche ein kontinuierliches Durchziehen der Substrate durch die Herstellungsanlage erlaubt, sehr gut für organische Solarzellen geeignet. Zudem zeichnen sich organische Halbleiter durch hohe Absorptionskoeffizienten aus, die 1000 mal besser sind als bei indirekten Halbleitern wie Silizium. Daher reichen sehr dünne Schichten von wenigen 100 Nanometern, um praktisch alle Photonen zu absorbieren, welche innerhalb der Absorptionsbandbreite liegen. Derzeit werden zwei Ansätze verfolgt, die sich auf die benutzte Materialklasse beziehen: Moleküle oder Polymere. Erstere werden meist im Vakuumprozeß thermisch verdampft, letztere werden aus einer Lösung abgeschieden. In den letzten Jahren haben verstärkte Forschungsanstrengungen zu beachtlichen Fortschritten geführt, nichtsdestotrotz sind organische Solarzellen derzeit weder in Hinblick auf die Effizienz — erreicht wurden bisher ~5% Wirkungsgrad mit beiden Ansätzen — noch auf die Lebensdauer mit anorganischen Solarzellen konkurrenzfähig. Bei der Effizienz wird man voraussichtlich auch mittelfristig unterhalb der von anorganischen Hochleistungszellen bleiben, dies kann jedoch durch die finanziell und energetisch günstigen Herstellungskosten und die flexiblere Einsetzbarkeit wettgemacht werden. Mit anderen Worten: organische Solarzellen sollen die derzeitigen Technologien nicht ersetzen, aber haben das Potential, sie zu ergänzen, und neue Anwendungsgebiete zu erschliessen.

## Physikalische Grundlagen

In einer klassischen anorganischen Solarzelle werden durch absorbiertes Licht Ladungsträgerpaare erzeugt, je ein Elektron und ein Loch, welche nur schwach aneinander gebunden sind. Durch den Einfluß eines p-n Überganges und des damit verbundenen Potentialgefälles innerhalb der Solarzelle werden die Elektron–Loch Paare getrennt und zu den jeweiligen Kontakten transportiert. In organische Halbleitern hingegen sind die Rahmenbedingungen anders [1]. Die Abschirmlänge ist wesentlich größer als in anorganischen Halbleitern, was zu einer stärkeren Interaktion der positiven und negativen Ladungsträger miteinander führt. Daher ist die primäre optische Anregung auch exzitonischer Natur, also ein stark gebundenes Elektron–Loch Paar. Zudem sind für technische Anwendungen meist nicht Einkristalle relevant, sondern eher amorphe und polykristalline Strukturen. Der elektrische Transport findet aufgrund der fehlenden langreichweitigen Ordnung nicht als Bewegung eines quasi-freien Ladungsträgers im Band statt, sondern mittels durch Hüpfen von einem lokalisierten Zustand zum nächsten [2]. Diese Rahmenbedingungen haben Kon-sequenzen auf die Funktionsweise der organischen Solarzellen und auf deren konzeptionelle Realisierung. Die ersten organischen Solarzellen wurden aus einem einzelnen Material hergestellt. Bei der Absorption von Licht werden aufgrund der geringen Abschirmung stark gebundene Coulomb-gebundene Elektron–Loch Paare, sogenannte Singulett-Exzitonen, erzeugt. Diese müssen getrennt werden, um letztlich einen Photostrom erhalten zu können. Um die Bindungsenergie zu überwinden, ist man entweder auf die thermische Energie angewiesen, oder eine Ladungstrennung an den Kontakten. Die Effizienz beider Prozesse stellte sich als sehr gering heraus: Die Temperatur ist im normalen Betrieb nicht hoch genug, und die Probendicke ist viel größer als die Exzitonendiffusionslänge. Die Konsequenz: die Exzitonen werden nicht getrennt, sondern rekombinieren. Dies führt meist zu Lumineszenz — und leuchtende Solarzellen sind nicht besonders effizient, da kein nennenswerter Strom generiert wird. Erst die Einführung von Zweilagenschichten Mitte der 80er Jahre brachte eine Besserung [3]. Die in einem der beiden Materialien, beispielsweise dem sogenannten Donator, absorbierten Lichtquanten erzeugen Exzitonen, welche zu der Grenzfläche beider Materialen diffundieren. Das andere Material sollte dann stark elektronenanziehend sein, und wird daher Akzeptor genannt; ein prominentes Beispiel ist das Buckminsterfulleren (C60). Die Energiedifferenz zwischen dem Elektronenniveau des Donators und des entsprechenden Akzeptorniveaus muß dabei größer sein als die Bindungsenergie des Exzitons. Bewegt sich also ein Exziton — durch Diffusion, da es keine

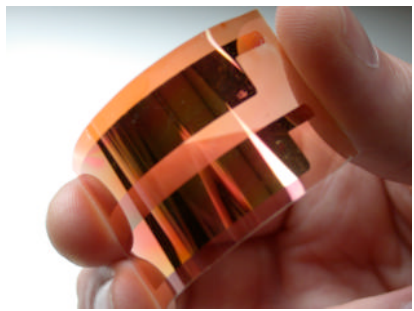

Abb. 1: *Organische Polymer–Fulleren Solarzelle auf flexiblem Substrat.*

## DIE SOLARZELLE

Die wichtigsten Kennzahlen, die die Leistungsfähigkeit einer Solarzelle beschreiben, sind die Leerlaufspannung, der Kurzschlußstrom, der Füllfaktor, und der Wirkungsgrad. Der Füllfaktor ist gegeben durch den Quotienten aus der maximalen Leistung (gelbes Rechteck) und dem Produkt aus Leerlaufspannung und Kurzschlußstrom (weißes Rechteck); er beschreibt somit die Rechteckigkeit der Strom–Spannungs-Charakteristik. Der Wirkungsgrad ist der Quotient aus maximaler Leistung und eingestrahlter Lichtleistung. Bei organischen Solarzellen ist der Photostrom wegen der feldunterstützten Ladungsträgertrennung abhängig von der Spannung: Daher wird der maximale Photostrom oftmals erst bei negativen Spannungen erreicht; dies verkleinert den Füllfaktor und den Kurzschlußstrom. Die Strom–Spannungs-Charakteristik anorganischer Solarzellen wird oftmals mit der Shockley-Gleichung beschrieben. Eine positive Spannung führt dabei zu der Injektion von Ladungsträgern in die Solarzelle, welche die Eigenschaften einer Diode aufweist. Im Idealfall hängt der Strom dann exponentiell von der Spannung ab. In einer realen Solarzelle kommt es jedoch zu Verlusten, welche durch die Erweiterung der Shockley-Gleichung um zwei Widerstände berücksichtigt werden können. Der sogenannte „Serienwiderstand" — in Serie zur Diode geschaltet — beschreibt unter anderem Kontaktwiderstände wie Injektionsbarrieren und Flächenwiderstände. Der „Parallelwiderstand" hingegen beschreibt den Einfluß lokaler Kurzschlüsse der beiden Elektroden, also Strompfade, welche an der Diode vorbei führen. Bei der Beschreibung organischer Solarzellen bekommt man jedoch Probleme: der „Parallelwiderstand" ist plötzlich abhängig von der Spannung und der Lichtleistung, der „Serienwiderstand" ist ebenfalls spannungsabhängig. Hell- und Dunkelstrom schneiden sich, d.h. daß der Photostrom spannungsabhängig ist. Während es bisher noch keine analytische Gleichung gibt, die die Strom–Spannungs-Charakteristik organischer Solarzellen beschreibt, so kennt man doch die meisten Ursachen für die beschriebenen Unterschiede. Da organische Materialien nicht so leitfähig sind wie anorganische, bilden sich bei hohen Spannungen Raumladungszonen, welche zu einem scheinbar spannungsabhängigen „Serienwiderstand" führen. Der feldabhängige Photostrom wird im Text diskutiert. Zudem können auch hier die auf Inseln gefangenen Ladungsträger eine Raumladungszone bewirken. Beides führt zu dem scheinbar licht- und spannungsabhängigen „Parallelwiderstand". Und Kurzschlüsse bzw. Kontaktwiderstände gibt es natürlich auch in organischen Solarzellen.

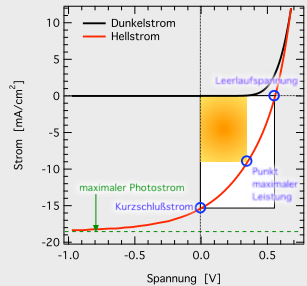
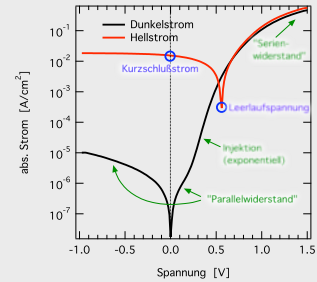

Nettoladung hat — an einen solchen Donator–Akzeptor Heteroübergang, ist es energetisch günstiger, wenn das Elektron auf das Akzeptormolekül übergeht. Dieser Trennungsprozess ist sehr schnell (an Polymer–Fulleren Grenzflächen schneller als 100 Femtosekunden) und daher effizient, weil alternative Verlustprozesse deutlich langsamer vonstatten gehen [4]. Bei dem sogenannten Elektronentransfer verbleibt das Loch auf dem Polymer: die Ladungsträger sind getrennt. Obwohl jetzt auf verschiedenen Materialien befindlich, sind beide noch immer über die Coulombkraft aneinander gebunden und somit lokalisiert, rekombinieren aber nicht mehr so schnell wie ein Exziton. Ein weiterer Schritt ist für die letztendliche Trennung vonnöten, und dieser erfordert ein elektrisches Feld — diese Abhängigkeit manifestiert später sich in einem für organische Solarzellen typischen feldabhängigen Photostrom, welcher sowohl den Füllfaktor als auch den Kurzschlußstrom der Solarzellen beinflußt. Wenn kein oder nur ein niedriges elektrisches Feld anliegt, ist die sogenannte monomolekulare Rekombination des Ladungsträgerpaares sehr wahrscheinlich. Nur wenn die feldunterstützte Ladungsträgerpaartrennung erfolgreich ist, können sowohl Elektron als auch Loch zu den jeweiligen Elektroden hüpfen, um dort als Photostrom gemessen zu werden. C. W. Tang, der diese Solarzelle 1986 mittels zweier konjugierter Moleküle realisierte, erreichte einen Wirkungsgrad von immerhin einem Prozent. Limitierend ist, daß die für eine vollständige Absorption des einfallenden Lichts benötigte Schichtdicke (~100nm) viel größer als die Diffusionslänge der Exzitonen (~10nm) ist. Meist liegt die Diffusionslänge also deutlich unterhalb der Absorptionslänge, so daß das Potential der Zweilagen-Solarzellen nicht ausschöpft wird.

Anfang der 90er Jahre wurde ein neues Konzept vorgestellt, welches sowohl der geringen Exzitonendiffusionslänge als auch der benötigen Schichtdicke Rechnung trägt: die in Fachkreisen *bulk heterojunction solar cell* bezeichnete Solarzelle mit einem sogenannten verteilten Heteroübergang [5] — das bedeutet, daß sich Donator- und Akzeptormaterial gegenseitig durchdringen, und somit deren Grenzfläche nicht mehr zweidimensional ist, sondern räumlich verteilt. Dieses Konzept wurde ursprünglich mittels eines aufgeschleuderten Polymer–Fulleren Gemisches realisiert, läßt sich aber auch bei Solarzellen aus konjugierten Molekülen mittels Koverdampfung erhalten. Verteilte Grenzflächen haben den Vorteil, daß Exzitonen über die gesamte Ausdehnung der Solarzelle hinweg sehr effizient getrennt, und Ladungsträger generiert werden können. Nachteil ist jedoch, daß die getrennten Ladungsträger wegen der höheren Unordnung langsamer transportiert werden. Zudem können sie in tiefen Störstellen oder Materialbereichen gefangen sein, welche nicht durch Perkolation mit der entsprechenden Elektrode verbunden sind, und so mit mobilen Ladungsträgern rekombinieren. Die wichtigsten Prozesse sind schematisch in Abb. 2 dargestellt.

Für eine effiziente Solarzelle mit verteilter Grenzfläche ist also eine gute Kontrolle der Morphologie sehr wichtig. Vergleichsweise einfache Möglichkeiten der Optimierung wurde dabei erst im neuen Jahrtausend erfolgreich angewandt. Die Wahl des Lösungsmittels [6] sowie das Tempern von flüssig prozessierten Polymer–Fulleren Solarzellen [7] führten je zu einer günstigeren inneren Struktur sowohl für die Ladungsträgertrennung, als auch den Transport. So wurde eine Vervielfachung des Wirkungsgrades erreicht, im Falle des Temperns von einem halben auf über 3%. Inzwischen hat man die Solarzellen mit den verteilten Grenzflächen weiter verbessert; bei aufgedampften Kupferphthalozyanin/Fulleren Zellen sind 5,0% Wirkungsgrad erreicht worden [8], bei flüssig prozessierten Polythiophen–Fulleren Zellen sogar 5,8% [9].

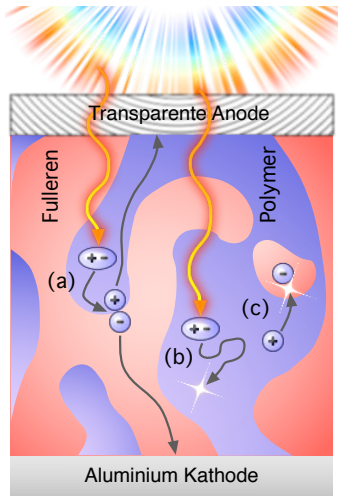

Abb. 2: *Ladungsgeneration in einer Polymer–Fulleren Solarzelle (nicht maßstabsgetreu). (a) Ein absorbiertes Photon erzeugt ein Exziton im Polymer, welches an die Grenzfläche diffundiert. Das Elektron wird auf das Fulleren transferiert, das resultierende Ladungsträgerpaar ist noch Coulombgebunden. Nach (hoffentlich) erfolgreicher feldunterstützter Trennung können Elektron und Loch zu den jeweiligen Elektroden hüpfen. (b) Das generierte Exziton zerfällt, da es innerhalb seiner Diffusionslänge keine Grenzfläche erreicht hat. (c) Ein photogeneriertes, schon freies Loch driftet Richtung Anode, trifft aber auf ein gefangenes Elektron und rekombiniert.*

### Neue Konzepte

Um organische Solarzellen weiter zu verbessern, müssen neue Donator- und Akzeptormaterialien synthetisiert werden, die neben der Fähigkeit zur Selbstorganisation — wichtig für eine hohe Ordnung der resultierenden Schichten — ein möglichst breites Absorptionsspektrum bieten, um das Sonnenlicht besser als bisher nutzen zu können. Bisher wird meist nur das Licht im Donatormaterial effizient absorbiert, es gibt also hier ein großes Potential zur Erhöhung des Photostromes. Zudem kann durch die Variation der relativen Energieniveaus von Donator- und Akzeptormaterial der Energieverlust beim Elektronentransfer verringert werden; dies wirkt sich direkt positiv auf die Leerlaufspannung aus. Aber auch auf Basis der bestehenden Zellen kann mittels eines schon von den anorganischen Solarzellen bekannten Konzepts ein erhöhter Wirkungsgrad erzielt werden: mit sogenannten Tandem-Solarzellen. Dabei werden zwei oder mehr Solarzellen mit sich ergänzenden Absorptionsbereichen und möglichst ähnlichem Kurzschlussstrom übereinander prozessiert; die Leerlaufspannung addiert sich dann auf. So erhält man auf gleicher Fläche eine Solarzelle mit deutlich höherem Wirkungsgrad. Bei diesem Konzept gibt es noch viel Optimierungspotential, aber die Anfänge sind gemacht: Solarzellen, aus immerhin sechs Einzelzellen bestehend, sind schon vorgestellt worden. Vorteilhaft ist allerdings, daß die Schichten zwischen den Einzelzellen auch aus der Flüssigphase, und somit kostengünstig, prozessiert werden können. Eine Abschätzung der maximalen Effizienz organischer Tandem-Solarzellen ist in Abb. 3 gezeigt. Weiterhin gibt es sogenannte Hybridsolarzellen, bei denen — ähnlich der Farbstoffsolarzelle — ein organisches Donatormaterial mit einem anorganischen Akzeptor genutzt werden. Der Akzeptor ist dabei nanoporös oder besteht aus Nanoteilchen, z.B. CdSe [11] oder ZnO, und soll somit für eine günstige Morphologie für Ladungsgeneration und -transport bieten. So können die Vorteile von organischen und anorganischen Materialien vorteilhaft verbunden werden.

### Marktpotential und Ausblick

Organische Solarmodule kann man derzeit nicht käuflich erwerben. Als Eingangsschwelle für kommerzielle Spartenanwendungen werden 10% Wirkungsgrad für einzelne Zellen, 5% für Module, und 3–5 Jahre Lebensdauer genannt [12]; diese Kriterien sind bisher noch nicht erreicht. Eine höhere Lebensdauer kann intrinsisch über neu designte Materialien, als auch extrinsisch mittels geeigneter — teils flexibler — Verkapselung erreicht werden. Dabei kann teilweise auf die Erfahrungen aus der OLED-Industrie zurückgegriffen werden. Wege zu höheren Wirkungsgraden sind oben schon beschrieben worden. Der Einstieg bei schon 5% Modulwirkungsgrad wird erst ermöglicht durch die voraussichtlich günstigen Herstellungskosten der organischen Solarzellen. Die Aufbringung durch Tintenstrahldrucker oder Niedrigvakuumverfahren, wie die sogenannte *organic vapor phase deposition*, erlauben eine Strukturierung der einzelnen Schichten während der Materialabscheidung, sowie eine homogene Abscheidung auf großen Flächen. Dadurch, daß der Produktionsprozess nicht für Lithographieschritte unterbrochen werden muß, sind höhere Produktionsvolumina möglich.

Insgesamt konnten in Hinblick auf die eingangs erwähnte Vision der günstigen, effizienten und flexiblen Solarzellen aus organischen Materialien schon beachtliche Fortschritte erzielt werden. Bis es zum kommerziellen Produkt kommt, sind allerdings noch weitere interdisziplinäre Anstrengungen von Physik und Chemie, sowie Verbesserungen der Produktionstechnologie notwendig, um das grundlegende Verständnis der organischen Solarzelle zu erlangen, und somit die Voraussetzung für Optimierungen zu schaffen.

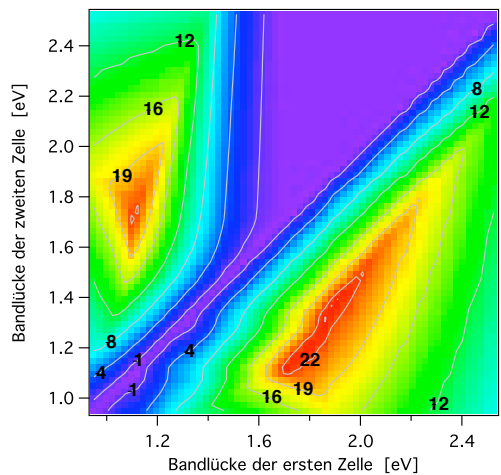

Abb. 3: *Eine Abschätzung des Wirkungsgrades (fett gedruckt, in Prozent) für eine Tandemzelle aus zwei übereinandergeschalteten Polymer–Fulleren Solarzellen mit unterschiedlichen Bandlücken. Eine solche Abschätzung ist keine Maximalabschätzung, da im Vergleich zu den bekannten Abschätzungen von anorganischen Solarzellen — Wirkungsgrad versus Bandlücke — wesentlich mehr Parameter eingestellt werden müssen. Hier wird u.a. angenommen, daß beide Solarzellen jeweils nur in einem 250nm breiten Absorptionsband Ladungsträger generieren können. Das absorbierte Licht wird direkt in einen Photostrom umgerechnet. Die Leerlaufspannung wird mit dem Modell von Koster et al. [10] abgeschätzt. Die Exzitonenbindungsenergie ist mit 300meV angesetzt worden. Der Füllfaktor, analytisch bisher nicht bestimmt, wurde mit 80% großzügig abgeschätzt. Der Strom der beiden Zellen wird der Einfachheit halber als min(Strom1, Strom2) angesetzt, die Spannungen werden aufaddiert.*